\documentclass[3p,twocolumn]{elsarticle}

\usepackage{amssymb}
\usepackage{amsmath}
\usepackage{lineno}
\journal{Chaos Solitons and Fractals}

\begin{document}
\begin{frontmatter}

\title{Breaking unidirectional invasions jeopardizes biodiversity in spatial May-Leonard systems}

\author[label1]{D. Bazeia}
\author[label2]{B.F. de Oliveira}
\author[label2]{J.V.O. Silva}
\author[label3]{A. Szolnoki}

\address[label1]{Departamento de F\'\i sica, Universidade Federal da Para\'\i ba, 58051-970 Jo\~ao Pessoa, PB, Brazil}

\address[label2]{Departamento de F\'\i sica, Universidade Estadual de Maring\'a, 87020-900 Maring\'a, PR, Brazil}

\address[label3]{Institute of Technical Physics and Materials Science, Centre for Energy Research, Hungarian Academy of Sciences, P.O. Box 49, H-1525 Budapest, Hungary}

\begin{abstract}
Non-transitive dominance and the resulting cyclic loop of three or more competing species provide a fundamental mechanism to explain biodiversity in biological and ecological systems. Both Lotka-Volterra and May-Leonard type model approaches agree that heterogeneity of invasion rates within this loop does not hazard the coexistence of competing species. While the resulting abundances of species become heterogeneous, the species who has the smallest invasion power benefits the most from unequal invasions. Nevertheless, the effective invasion rate in a predator and prey interaction can also be modified by breaking the direction of dominance and allowing reversed invasion with a smaller probability. While this alteration has no particular consequence on the behavior within the framework of Lotka-Volterra models, the reactions of May-Leonard systems are highly different. In the latter case, not just the mentioned ``survival of the weakest'' effect vanishes, but also the coexistence of the loop cannot be maintained if the reversed invasion exceeds a threshold value. Interestingly, the extinction to a uniform state is characterized by a non-monotonous probability function. While the presence of reversed invasion does not fully diminish the evolutionary advantage of the original predator species, but this weakened effective invasion rate helps the related prey species to collect larger initial area for the final battle between them. The competition of these processes determines the likelihood in which uniform state the system terminates. 
\end{abstract}

\end{frontmatter}

\section{INTRODUCTION}

Preserving biodiversity has a paramount importance in all ecosystems especially nowadays when climate change causes rapidly altering living environments for species and their adaptations to the new conditions are hardly predictable \cite{oneill_s02,pacheco_plrev14,wang_x_axv20,szolnoki_epl17}. In general, biodiversity can be considered as a delicate balance between different processes including speciation, extinction, migration and others. Several theoretical theories have been suggested to understand its origin, and a surprisingly simple and powerful tool is offered by non-transitive dominance among competing members \cite{2018-Dobramysl-JPA-51-063001,szolnoki_jrsif14,park_c18c}. The latter situation is modeled by the well-known rock-scissors-paper game where every participant dominates another one and is dominated simultaneously by a third one \cite{2014-Avelino-PRE-89-042710,nagatani_em17,szolnoki_srep16b,guo_jrsif20}.

The two mainstream microscopic mathematical models which capture the essence of these relations are the so-called Lotka-Volterra ($LV$) and May-Leonard ($ML$) systems \cite{groselj_pre15,2019-Brown-PRE-99-062116,2012-Avelino-PRE-86-036112,nagatani_pa19b,2013-Roman-PRE-87-032148,2018-Bazeia-EPL-124-68001}. While in the former $LV$ approach the particle conservation is maintained because a dominant species occupies the empty space of dominated species immediately, there is no such constraint in $ML$ models. In the latter case the invasion is split into a selection and a probabilistic reproduction step which makes the sum of all species a non-conserved quantity.

Previous works highlighted that behaviors of the spatial cyclic dominant systems are remarkably robust with respect to model variations and both $LV$ and $ML$ models predict some universal features \cite{west_pre18,nagatani_jtb19,2017-Bazeia-SR-7-44900}. One of these is when varying reaction rates have little effect on the dynamical evolution \cite{2010-He-PRE-82-051909}, or quenched spatial disorder has only minor effect on species coexistence \cite{2011-He-EPJB-82-97}. A particularly interesting observation is the so-called ``survival of the weakest'' paradox which emerges when the invasion rates within the loop are heterogeneous. Counter-intuitively, in this case the ``weakest species'', who has the smallest invasion power, gains more and has the largest population in the stationary state \cite{tainaka_pla95, 2001-Frean-PRSLB-268-1323}. It was recently demonstrated that 
despite the different population dynamics and spatial patterns, both $LV$ and $ML$ formulations lead to qualitatively similar results and confirm the robustness of this effect \cite{2019-Avelino-PRE-100-042209}.

On the other hand, mean-field calculations warn us that the details of microscopic interactions between competing species can be important and careful studies are necessary to explore the frontiers of robustness. A well-known example is when we leave cyclic $LV$ model by separating selection and reproduction processes the resulting $ML$ model modifies the nonlinear dynamics from neutral orbits of $LV$ model to an unstable spiral \cite{2010-Frey-PA-389-4265}.

In our present work, motivated by the possible importance of microscopical details, we explore how the breaking of unidirectional invasions influences the evolutionary outcome. When we break the direction of dominance between a predator and a prey species and allow a reversed process with a certain probability then it conceptually may result in similar effect as was observed for heterogeneous invasion rates. Namely, the simultaneous usage of direct and reversed invasions will result in a decreased effective invasion rate, hence a weakened power of predator species. This picture is confirmed by a previous study of a $LV$ model where direct invasion was applied with probability $1-q$ and the related indirect process was executed with probability $q$ \cite{tainaka_pla95}. Such an intervention to the original model does not change the coexistence of species because all three species survive for all $q<0.5$ values, but the above mentioned ``survival of the weakest'' effect can be still observed.

The situation, however, is strikingly different when invasion in the reversed direction is allowed in the framework of $ML$ systems. As we will demonstrate, here not just the coexistence of species, hence biodiversity, is jeopardized, but also ``the survival of the weakest'' effect vanishes. These observations highlight that we should be careful when we estimate the robustness of some effect based solely on a single theoretical approach. 

Our paper is organized as follows: in the next Sec. \ref{sec:model} we describe the model details, the applied microscopic rules and specify how extinction to a homogeneous state is evaluated. In Sec. \ref{sec:result} we present our main results and provide detailed explanation to the observed behaviors. Finally, in Sec. \ref{sec:discussion} we conclude with some discussion and potential issues for future investigations.

\section{BREAKING UNIDIRECTIONAL INVASIONS}
\label{sec:model}

In the present work, we start from the classical spatial rock-scissors-paper game where three species, red ``1'', blue ``2'', and green ``3'' dominate  each other. The lattice sites are occupied by one of these species, or remain empty which is marked by ``0''. According to the $ML$ approach the microscopic rules contain a mobility step (with probability $m$), a competition or predation (with probability $p$), and a reproduction step (with probability $r$), where the $m+p+r=1$ constraint is used \cite{2014-Avelino-PRE-89-042710}. In most cases we apply $m=0.50$, $r=0.25$ and $p=0.25$, but other values of parameters are also considered, as specified later.

Our numerical simulations are always started from random initial conditions, where competing species and empty sites occupy distinct positions with equal probability. We apply $L \times L$ square lattice with periodic boundary conditions. In each time step a single individual (active) is chosen randomly to interact with a randomly selected neighbor (passive). One generation is the unity of time that corresponds to  $L^2$ successful interactions. When mobility is selected, the active and passive neighbors exchange their positions. Reproduction occurs when the neighboring passive site is empty. In this case, the empty neighbor is colored with the same color of the active site. Last, when competition is considered, then it follows the usual cyclic rules. Namely, if the active site is occupied by species $i$ and the passive by $i+1$ in cyclic manner then the passive site becomes empty and changes its color to white.

As a technical note, during the simulations we have used linear system size from $L=50$ to $L=500$ to reveal the possible finite-size effects. The applied observation time was up to 50000 full generations, and to reach the requested accuracy we have averaged the results of 1000 independent runs.

\begin{figure}[h!]
\centering
\includegraphics[width=7.8cm]{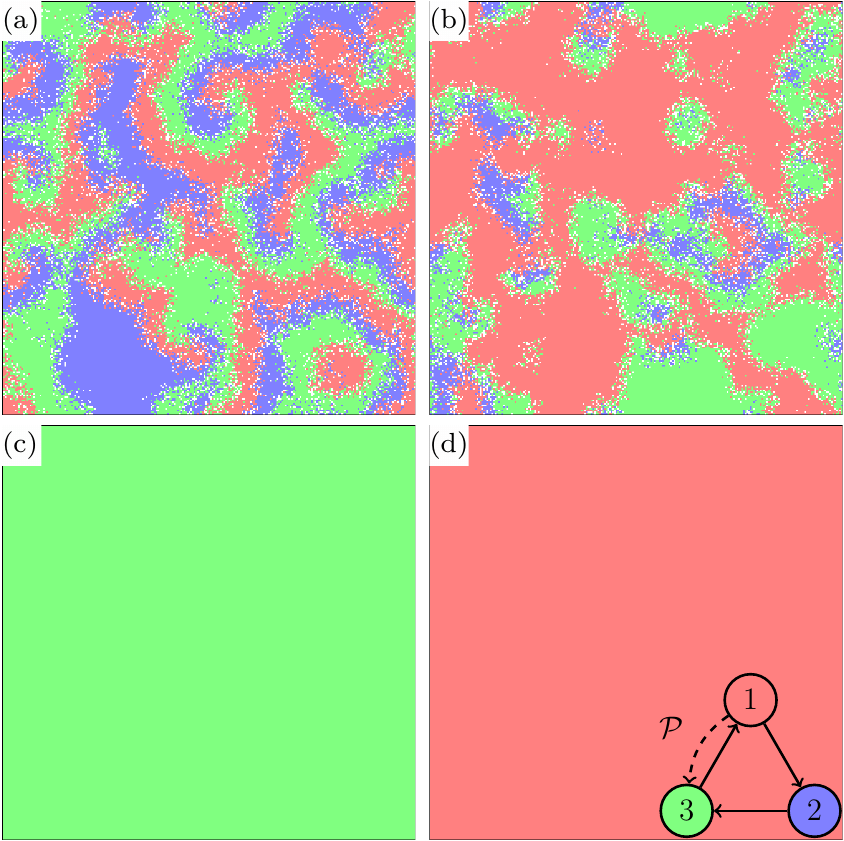}\\
\caption{Snapshots of our $ML$ system illustrating the final destinations for $\mathcal{P}=0$ (a), $\mathcal{P}=0.25$ (b), $\mathcal{P}=0.50$ (c) and  $\mathcal{P}=0.75$ (d), after 5000 generations on a $L \times L = 250 \times 250$ grid. While for small $\mathcal{P}$ values the coexistence is preserved, for large $\mathcal{P}$ values the system terminates onto a homogeneous state. Inset in panel~(d) depicts the food-web of our system, where solid arrows represent dominance with probability $p$, dashed arrow marks a reversed invasion with a probability $p \cal P$, where ${\cal P}\in [0,1]$. Other parameters are $m=0.5, p=0.25$, and $r=0.25.$}\label{snapshots}
\end{figure}

Beside the usual invasions we allow species ``1'' to invade species ``3'' with probability  $p \mathcal{P}$, where $\mathcal{P} \le 1$ serves a scaling factor of predation in the reversed order between species ``3'' and ``1''. In this way we not just weaken the invasion power of predator ``3'' toward prey ``1'', but at the same time we break the unidirectional flow of invasion. The summary of our extended model is plotted in the inset of panel~(d) of Fig.~\ref{snapshots}. Evidently, for $\mathcal{P}=0$ the model evolves according to the standard rock-scissor-paper system where rotating spiral patterns ensure the stable coexistence of competing species. When $\mathcal{P}=1$ then the dominance of species ``3'' over species ``1'' is absent, hence species ``3'' will die out due to the presence of predator ``2'' species. The latter, however, is vulnerable to species ``1'' who will gradually prevail in the whole space. Notably, when $\mathcal{P} < 1$, then there is a net flow around the loop which maintains biodiversity in $LV$ systems \cite{tainaka_pla95}. In this way $\mathcal{P}$ offers a control parameter to tune finely the strength of net invasion flow in the loop and also serves a link between the states of coexistence and homogeneous states which are present at the edges of $\mathcal{P}$ interval. 

\section{RESULTS}
\label{sec:result}

Our first key observations are summarized in Fig.~\ref{snapshots}, where we plot the snapshots of spatial distribution of species for some representative $\mathcal{P}$ values. At $\mathcal{P}=0$, shown in panel~(a), we detect the well-known rotating spiral pattern of standard model. For a moderate value of $\mathcal{P}$, shown in panel~(b), the pattern of the stationary state changes significantly. The above mentioned spirals vanish and larger homogeneous spots emerge, but the coexistence of competing species is still stable. Increasing $\mathcal{P}$ further, however, the biodiversity is lost and the system terminates mostly in a homogeneous state where species ``3'' prevails. This destination is shown in panel~(c). Finally, when $\mathcal{P}$ is high enough, the typical destination of evolution changes, as panel~(d) shows, and the system terminates exclusively into the state where only species ``1'' is present. 

\begin{figure}[h!]
\centering
\includegraphics[width=7.8cm]{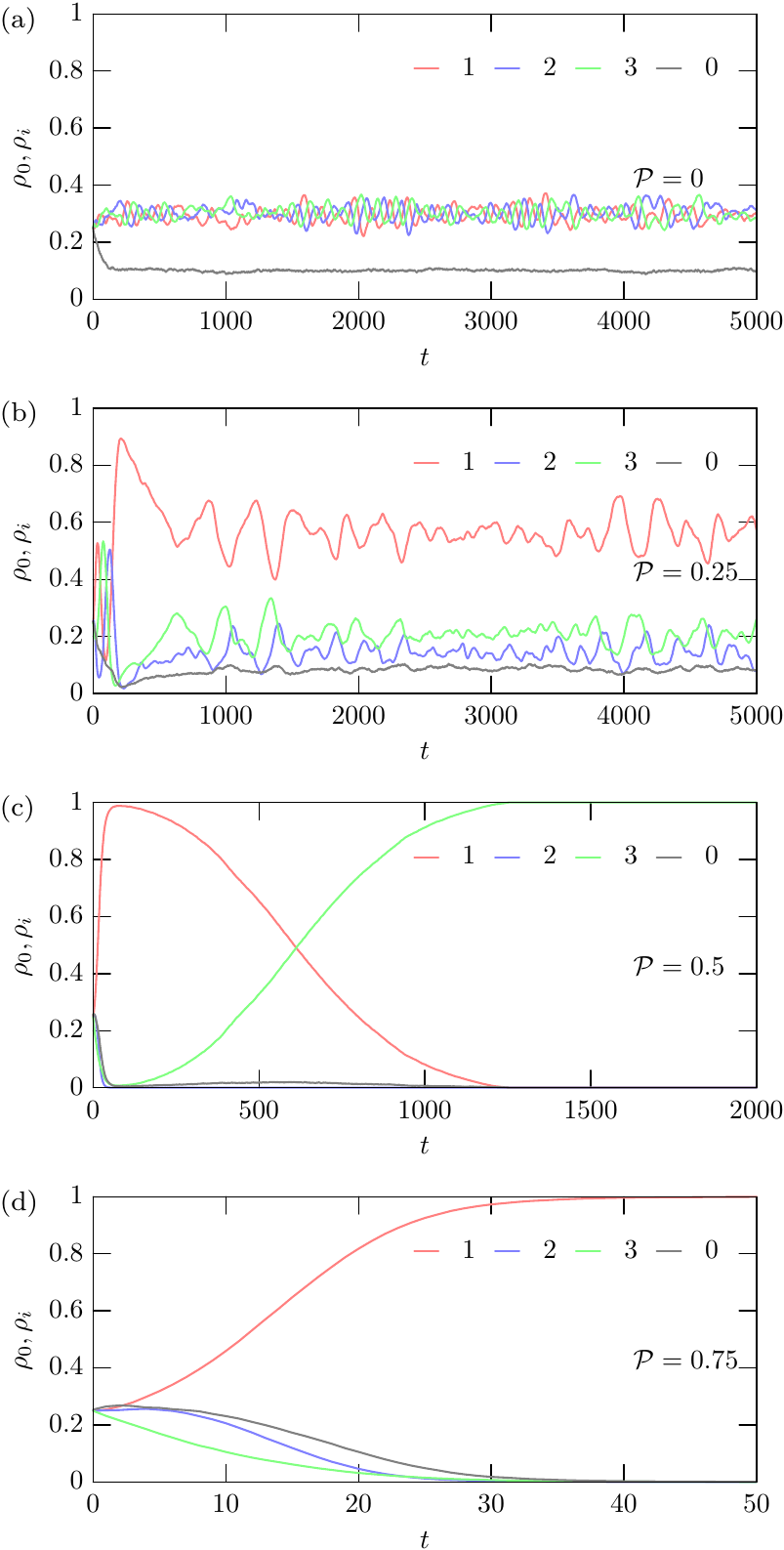}\\
\caption{Representative time evolution of microscopic states for $\mathcal{P}=0, 0.25, 0.5$ and $0.75$ from panel~(a) to (d) obtained at $L=250$ system size. Here ``1'' to ``3'' denote the competing species, while ``0'' depicts empty sites.}\label{time}
\end{figure}

To get further qualitative impressions about the time course of evolution for different $\mathcal{P}$ values, in Fig.~\ref{time} we present time evolution of abundance of different microscopic states. In panel~(a) we can see the standard behavior of the classic model. Namely, the concentrations of all three species fluctuates slightly around the same $0.3$ value, while the stationary portion of empty sites is about $0.1$. Naturally, these values depend on the model parameters $m, r$, and $p$.

When we break the exclusive direction of invasions and allow reversed process then the equivalence of species disappear. This would be an expected reaction of the system due to heterogeneous effective invasion rates between predator-prey pairs, but the portion of ``weakened'' species ``3'' becomes smaller comparing to the homogeneous loop case obtained at $\mathcal{P}=0$. This behavior is against to the broadly valid ``survival of the weakest'' effect, which warns us that breaking the rotation symmetry of invasion loop has a different impact on the $ML$ system as observed for $LV$ models. It is also worth mentioning that albeit the trajectory of empty site concentration is almost the same as earlier, but the fluctuations of species concentrations are enhanced, which is an indirect consequence of the disappearance of stable rotating spirals characterizing the $\mathcal{P}=0$ limit.

For higher $\mathcal{P}$ values, shown in panel~(c), the trajectory of evolution is significantly different. Here, there is no fluctuation, but the system goes almost deterministically toward a homogeneous destination. At first sight it may be counter-intuitive that species ``3'', who has two predators but only one prey in the loop, will win the game, but monitoring the trajectories give us the key. At early stage of the evolution it seems to be a full red dominance of species ``1'', but eventually green species ``3'' takes over the leading role and sweeps out its competitor. Here, it is crucial that species ``2'' goes extinct very early followed by the direct battle of the remaining two species.

Last, when the value of $\mathcal{P}$ is high enough, shown in panel~(d), the trajectory of evolution differs again. Here the green and blue lines become zero almost simultaneously, hence species ``3'', represented by the green line, is unable to enforce its advantage over red species ``1'', because its predator species ``2'' is always present. One may argue that the final victory of species ``1'' is expected because it has two preys and only one predator in the loop, but we must stress that there is still a net nonzero invasion around the loop for every  $\mathcal{P} < 1$ value, hence there would be a good reason for maintaining biodiversity.

It is a common feature of panels~(c) and~(d) that empty sites vanish only when the population becomes uniform, because in this case, in the absence of selection, reproduction of surviving species will eventually fill all available space. As we can see, there are two conceptually different ways to destroy biodiversity and in the following we will discuss their origins and consequences in detail.

\begin{figure}[h!]
\centering
\includegraphics[width=7.8cm]{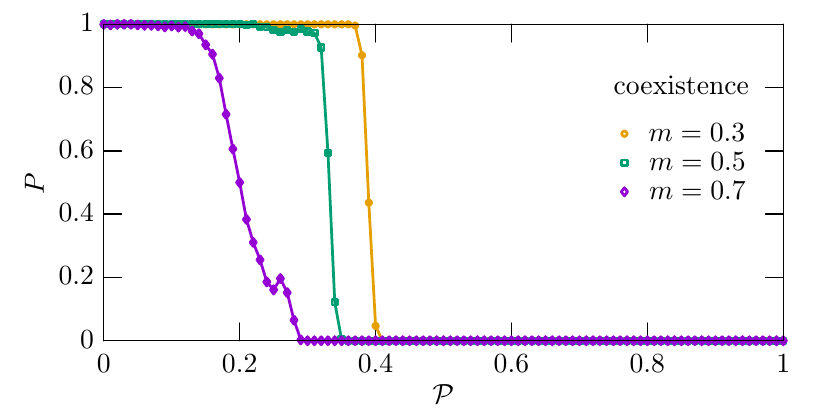}\\
\caption{Probability of maintaining coexistence in dependence of $\mathcal{P}$ for different values of mobility parameter as indicated in the legend. The system size are $L=250$ while the observation time is 20000 generations. Data for each $\mathcal{P}$ value were obtained by averaging over 1000 independent runs.}\label{coexistence}
\end{figure}

For a deeper insight we give quantitative description how biodiversity is lost as a consequence of breaking the unidirectional invasion rule. For this reason, by following a similar method applied in Refs.~\cite{2019-Avelino-PRE-100-042209,2007-Reichenbach-N-488-1046}, we run independent simulations and measured in how many cases the coexistence of three species survived after $t=20000$ generations. This quantity divided by the total number of runs determines the probability of coexistence at a given value of $\mathcal{P}$. The results are plotted in Fig.~\ref{coexistence} where we present the surviving probability in dependence of $\mathcal{P}$. Let us first focus on the green $m=0.5$ curve and discuss the mobility dependence later. Our first observation is the coexistence, hence biodiversity, is lost suddenly at a critical $\mathcal{P}$ value. When the strength of reversed invasion exceeds this threshold value then the coexistence of competing species cannot be maintained anymore and the system evolves to a uniform state where only one species is present. This behavior is in sharp contrast to the one observed for $LV$ models where allowing invasions in the reversed direction does not jeopardize the coexistence of species \cite{tainaka_pla95}. Naturally, as we noted, the fractions of competing species changes due to modified effective invasion rate between predator-prey interaction, but all of them survive.

\begin{figure}
\centering
\includegraphics[width=7.8cm]{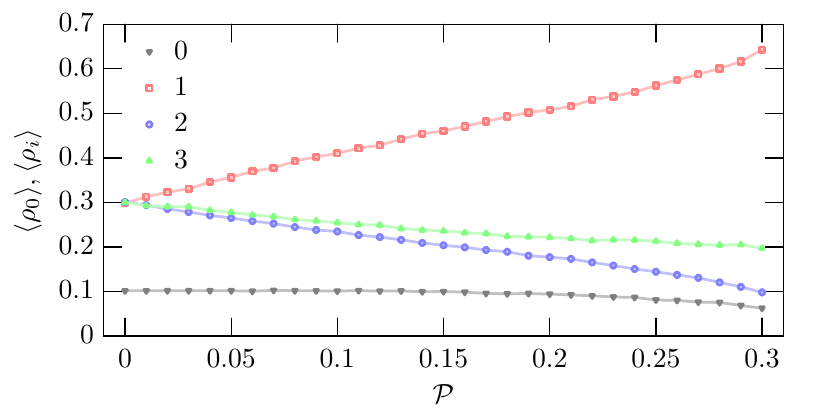}\\
\caption{The average abundance of species and empty sites in the stationary state as a function of $\mathcal{P}$ at $m=0.5$, $r=0.25$, $p=0.25$ values. Each point was averaged over 20000 generations where the first 5000 generations were excluded to measure the stationary state properly. Note that above $\mathcal{P} \simeq 0.34$ the coexistence cannot be maintained independently of the applied system size.}\label{weakest}
\end{figure}

Figure~\ref{coexistence} suggests that coexistence is maintained only for small values of $\mathcal{P}$. To explore this phase in detail we have measured the stationary fractions of species for different $\mathcal{P}$ values. The resulting plot of Fig.~\ref{weakest} highlights another deviation from the standard behavior of cyclically dominant systems. As we already argued, the introduction of invasions in the reversed direction weakens the dominance of species ``3'' over species ``1'' and this would imply the increment of the population of species ``3'' due to the survival of the weakest effect. Indeed, this happens exactly for $LV$ models \cite{tainaka_pla95}, but not in our present case. Here the concentration of species ``3'' decays gradually as we increase the strength of reversed invasion. Interestingly, the biggest loser is species ``2'' who is the predator of species ``3''. One may argue that the increment of the frequency of species ``1'' is not really surprising because the introduced reversed invasions start from species ``1''. However, in case of cyclically dominated systems such a simple explanation is not working, because we generally cannot reach the desired purpose by supporting a competitor directly. This is a broadly valid observation and emphasizes that we should be very careful when we want to control an ecological system.  

Next we explore how mobility influences the extinction caused by symmetry breaking of invasions. Earlier, within the framework of standard $ML$ model, it was already pointed out that mobility has a decisive role on biodiversity  \cite{2007-Reichenbach-N-488-1046,2018-Avelino-PRE-97-032415,mobilia_g16}. More precisely, with increasing weight of mobility, the well-known rotating spiral structures grow and the characteristic length of spiral arms increases. Above a threshold value of $m$ this length outgrows the system size and coexistence is replaced by a uniform population. When we discussed the consequence of symmetry breaking invasion on the emerging patterns, we already noted that this intervention also breaks the stable spirals and results in larger homogeneous spots as we increase $\mathcal{P}$. From these observations we may conclude that breaking the invasion loop has conceptually similar consequence on system evolution as the introduction of mobility has. Indeed, Fig.~\ref{coexistence}, where we plotted the coexistence probability for different values of mobility, confirms this argument nicely. In particular, by increasing $m$ the threshold value of $\mathcal{P}$ decreases. It suggests clearly that both mobility and the introduction of reversed invasion destroy biodiversity and these two effects support each other in a synergistic way.

Lastly, we focus on the parameter region where coexistence cannot be maintained and the system evolves into a homogeneous state where a single species occupies all available space. To quantify the possible destinations we launched simulations from independent random initial states and recorded the number of different final states. This quantity divided by the whole number of runs determines the survival probability of a specific species. Importantly, two different destinations were observed in the whole range of $0.34 \lesssim \mathcal{P}$ interval. Namely, the system evolved either into a state of full species ``3'' or into the uniform state of species ``1'', as we already noted earlier. 

\begin{figure}[h!]
\centering
\includegraphics[width=7.8cm]{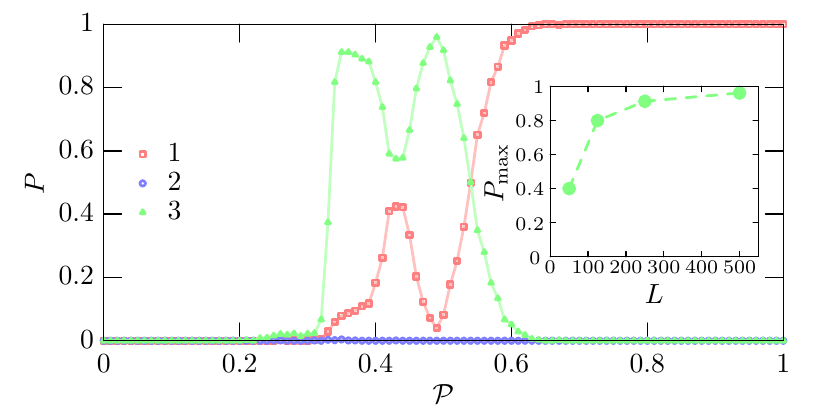}\\
\caption{Survival probability of a single specie in dependence of $\mathcal{P}$, which was calculated from 1000 independent runs at each $\mathcal{P}$ value. The inset shows the height of first peak for species ``3'' curve as a function of system size.}\label{survive}
\end{figure}

The survival probabilities depicted in Fig.~\ref{survive} highlight that the evolution to a uniform state is far from trivial. While in the $0.34 \lesssim \mathcal{P} \lesssim 0.55$ region the system evolves mostly into the full species ``3'' state, above $0.55 < \mathcal{P}$ the clear destination is full species ``1'' state. The dominance of species ``3'' in the intermediate $\mathcal{P}$ region is rather surprising because in the neighboring coexistence phase the portion of species ``1'' grows as we increase $\mathcal{P}$, therefore an analytical continuation would be the full species ``1'' solution. Still, the system evolves into an alternating destination in most cases. Further curiosity of the survival probability is its highly non-monotonous character. Namely, there are two peaks of green line characterizing the survival chance of species ``3'' before it becomes zero in the large $\mathcal{P}$ limit. Importantly, the destination to full green state is not a finite-size effect because this probability converges toward $1$ as we increase the system size. This finite-size analysis is shown in the inset of Fig.~\ref{survive}. 

To understand why species ``3'' has an optimal  $\mathcal{P}$ to survive we should consider some facts. An important observation is  species ``2'' always dies out first and leaves behind the battle between the remaining two species. Secondly, species ``3'' has an evolutionary advantage over species ``1'' because they are in predator-prey relation with $1- \mathcal{P}$ effective rate. Last, the above mentioned battle between species $1$ and $3$ starts from unbalanced conditions. More precisely, the starting population of species ``1'' always exceeds the fraction of species ``3'' at the moment when species ``2'' dies out.

Because of the favorable starting position of species ``1'' the battle would end very soon, unless species ``3'' has a chance to validate its evolutionary advantage. This case is shown in panel~(c) of Fig.~\ref{time}, where the system nearly terminates into the full red state, but before it green species gradually invades the whole space. This chance is higher at a larger system where the likelihood of surviving of a small domain occupied by species ``3'' is higher. This explains why the height of the green peak of survival probability is higher for larger system size, as shown in the inset of Fig.~\ref{survive}. For higher $\mathcal{P}$, however, the evolutionary advantage of species ``3'' becomes smaller, and in parallel the starting position for species ``1'' is more favorable: during the first stage of the evolution, when species ``2'' is present, larger $\mathcal{P}$ ensures higher chance for species ``1'' to collect larger territory for the final battle. These two effects support each other, hence green species ``3'' simply has no chance to fight efficiently because the final battles actually ends before it started. This specific trajectory can be seen clearly in panel~(d) of Fig.~\ref{time} where the final destination is reached very quickly.
 
\begin{figure}[h!]
\centering
\includegraphics[width=7.8cm]{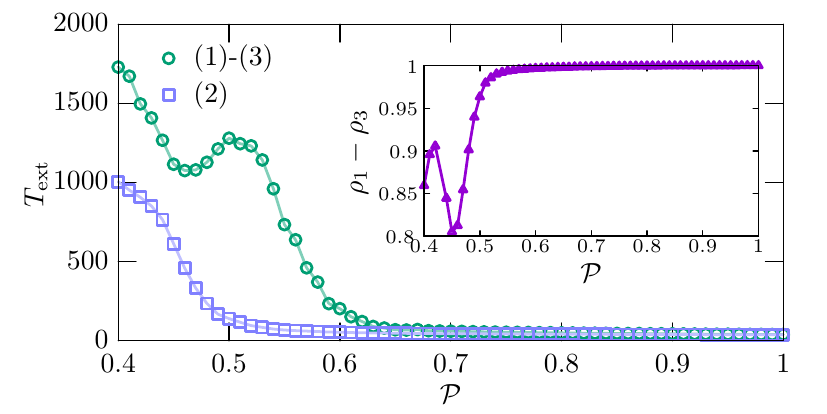}\\
\caption{Extinction times as a function of $\mathcal{P}$. Here $(2)$ marks the average time when specie ``2'' dies out. The curve, denoted by $(1)-(3)$, shows the average time when either species ``1'' or species ``3'' goes extinct hence biodiversity is lost. The results are averaged over 1000 independent runs at each $\mathcal{P}$ value. The inset shows the normalized excess of specie ``1'' population compared to species ``3'' at the very moment when species ``2'' dies out.}\label{extinction}
\end{figure}

But if we check the survival probability in Fig.~\ref{survive} then we can see a second peak in green line. To understand the origin of this unexpected behavior we also measured the extinction time of different species. Starting from a random initial state we monitored how the portions of species change and recorded the time when the system reached a uniform state. The applied method to calculate the average extinction time is similar to those used in Refs. \cite{2010-He-PRE-82-051909, 2011-He-EPJB-82-97, 2012-Dobrinevski-PRE-85-051903}. To analyze how survival probabilities vary in dependence of $\mathcal{P}$, it has a paramount importance that species ``2'' dies out first, therefore we recorded separately the average extinction time of this species. Our results are summarized in Fig.~\ref{extinction} where we plotted the above specified extinction times for $\mathcal{P}\in [0.4,1]$ interval where biodiversity cannot be maintained.

The first observation worth mentioning is the extinction time of species ``2'' is a monotonous function of $\mathcal{P}$. The higher $\mathcal{P}$, the sooner species ``2'' goes extinct. On the other hand, there is a local maximum in the final extinction time of the whole system at around $\mathcal{P} \approx 0.5$, as shown by the green line in Fig.~\ref{extinction}. This increment of extinction time signals a kind of ``frustration'' or ``hesitation'' about which destination to choose during the evolution. More precisely, the battle between the remaining two species seems less obvious, which implies a longer fight between them. Evidently, for higher $\mathcal{P}$ values we would expect  a clear, less ambiguous trajectory, still the chance to drift toward full species ``3'' state increases. This is because the starting condition for the final battle becomes less biased, hence more favorable for species ``3'' in the mentioned $\mathcal{P}$ interval. This effect is shown in the inset of Fig.~\ref{extinction} where we plotted the  excess population of species ``1'' compared to species ``3'' in a normalized way at the very moment when species ``2'' dies out. The local minimum of the excess population means an additional help for species ``3'' who will have a better chance to fight against species ``1''. This explains the local maximum of extinction time and the second maximum in survival probability of greens.

\section{DISCUSSION}
\label{sec:discussion}

Cyclic dominance is always the source of counter-intuitive phenomenon in population dynamics \cite{2016-Roman-JTB-403-10,szolnoki_csf20b,park_srep17,hashimoto_jpsp18,2018-Avelino-EPL-121-48003,baker_jtb20,nagatani_jpsj20,2019-Bazeia-PRE-99-052408,baker_jtb20,szolnoki_prx17}. Hence designing such ecosystems could be an intellectual challenge because naive intervention into these systems may result in undesired consequences \cite{park_c18,2018-Shadisadt-PRE-98-062105}. We should also note that the mentioned non-transitive relations are not restricted to ecological systems, but can occur in social dilemmas, too \cite{szolnoki_njp14,canova_jsp18,szolnoki_prl12,wang_x_rspa20,szolnoki_epl15,amaral_pre20,szolnoki_pre15,hauert_s02,yang_lh_epl18,szolnoki_njp18b}. Therefore to clarify the possible reactions of such systems could be a vital task.

Different theoretical approaches are used to resolve this job and their comparative analysis warns us that microscopic details may matter and we should be careful when we want to estimate the robustness of a specific observation. Motivated by these experiences in the present work we explored the possible consequences of breaking the direction of invasion flow in the framework of the May-Leonard system. The extension of the traditional model is tiny, but its dynamical consequences are astonishing.

We first demonstrated that biodiversity can be jeopardized by breaking unidirectional invasions. If the latter exceeds a threshold value then previously observed coexistence of species cannot be maintained anymore. This is in stark contrast to the system behavior observed for $LV$ models where similar reversed direction invasion is allowed \cite{tainaka_pla95}. In this way, beside intensive mobility, the chance of invasions in reversed direction offers an alternative mechanism to destroy biodiversity.

We also pointed out that the well-known and frequently studied survival of the weakest effect has limited validity. This phenomenon remains robust both in $LV$ and in $ML$ approaches when invasion rates are homogeneous, but they still point to the same clockwise rotation in the loop \cite{2019-Avelino-PRE-100-042209}. But allowing invasions anticlockwise breaks this peaceful picture. While the mentioned effect remains still valid in $LV$ systems \cite{tainaka_pla95}, but in $ML$ systems it does not survive anymore. In the latter case the ``weakened'' predator is literally weakened because its portion decreases gradually as we increase the strength of reversed invasion.

As we demonstrated, the extinction to a homogeneous state shows further interesting features. First, the above mentioned ``weakened'' species has a clear chance to win the final battle at moderate reversed invasion strength, which is against our preliminary expectations. When we increase the reversed invasion strength further the survival probability of the mentioned species decays, then increases again before becoming zero. The last decay is in agreement with the naive picture we built based on the food-web of the model, but the second optimum of survival is related to a frustration of the system when choosing between the available destinations. This frustration is quantified by measuring the extinction time to reach the homogeneous state and it is related to the additional support obtained by the weakened species in an intermedium interval of reversed invasion strength. Evidently, this extra support of weakened species via better starting condition for the last  battle is related to the subtle interaction how and when its predator species dies out, because the latter event has a decisive role on the winning chance of competing species.

\vspace{0.5cm}

D.B. acknowledges Conselho Nacional de Desenvolvimento Cient\'\i fico e Tecnol\'ogico (CNPq, Grants nos. 306614/2014-6 and 404913/2018-0) and Para\'\i ba State Research Foundation (FAPESQ-PB,  Grant no. 0015/2019) for financial support. B.F.O. and J.V.O.S. thank CAPES - Finance Code 001, Funda\c c\~ao Arauc\'aria, and INCT-FCx (CNPq/FAPESP) for financial and computational support. A.S. is grateful to the Hungarian National Research Fund (Grant K-120785).

\end{document}